# Investigating spin and orbital effects via spin-torque ferromagnetic resonance

J. L. Costa[1], E. Santos[1], A. Y. M. Tani[1], J. B. S. Mendes[2], and A. Azevedo*[1]

[1] *Departamento de Física, Universidade Federal de Pernambuco, Recife, Pernambuco 50670-901, Brazil.*
[2] *Departamento de Física, Universidade Federal de Viçosa, 36570-900 Viçosa, Minas Gerais, Brazil*

**Abstract**

In this work, we experimentally investigate spin and orbital torque phenomena using the spin-torque ferromagnetic resonance (ST-FMR) technique in a series of bilayer systems composed of different normal metal (NM) materials. Permalloy (Py) and Ni were employed as ferromagnetic (FM) layers to probe the spin and orbital torque responses, respectively. For the $SiO_2$/FM/NM bilayers, we extracted the damping-like and field-like torque components, as well as the damping-like torque efficiency for each sample, and compared our results with previously reported numerical and experimental data in the literature. Additionally, we experimentally demonstrate the presence of an out-of-plane torque component, which we attribute to interfacial mechanisms and associate with a spin-orbital polarized current along the *z*-direction. This interpretation is supported by the azimuthal angular dependence of the applied magnetic field. Our results provide compelling evidence of orbital torque associated with the orbital Hall effect in several materials, thereby broadening the prospects for magnetization switching driven by orbital torque.

*Corresponding author: antonio.azevedo@ufpe.br

## I. Introduction

The electrical control of magnetization in nanoscale devices relies on the conversion of charge currents into spin currents, which subsequently exert torques on the magnetization of a ferromagnetic layer. The two primary and long-established mechanisms for this are the spin-transfer torque (STT)[1-3] and the spin-orbit torque (SOT)[4,5]. While STT originates from the injection of a spin-polarized current from a second magnetic layer, SOT is generated within an adjacent non-magnetic (NM) layer with strong spin-orbit coupling. In a typical bilayer structure, such as NM/FM, the central physical effect is often the Spin Hall Effect (SHE)[6-10]: a charge current flowing in the NM layer is transversely converted into a pure spin current, which is then injected into the adjacent ferromagnet (FM) and exerts a torque on its magnetization. However, interfacial mechanisms, such as the Rashba-Edelstein effect (REE), may also contribute to the measured torques. This scheme provides an efficient route to electrically control and switch the magnetization using a single ferromagnetic layer. More recently, the field of orbitronics has revealed that the orbital degree of freedom can play an equally critical, and sometimes dominant, role[11-26]. Analogous to the SHE, the Orbital Hall Effect (OHE) in certain materials, mainly including light elements, can generate a perpendicular flow of orbital angular momentum (an orbital current) in response to an electric field. This orbital current can subsequently be converted into a spin current via spin-orbit coupling (SOC), either in the ferromagnet itself or at the interface, ultimately contributing to a magnetization torque. This indirect pathway defines the orbital torque (OT) mechanism. Consequently, in NM/FM heterostructures the measured torque can contain intertwined contributions from both the SHE and the OHE, mediated by their respective charge-to-spin and charge-to-orbital-to-spin conversion processes, as well as interfacial Rashba-type effects. While SOT has enabled efficient magnetization switching, recent advances in orbitronics suggest that orbital angular momentum, mediated via the OHE, can provide an alternative or complementary pathway for torque generation.

Despite their promise, investigations of STT and SOT based on direct current switching face significant experimental challenges, including low signal sensitivity, difficulty in clearly separating damping-like and field-like torque components, and the need for high current densities that induce undesirable heating and complex domain dynamics. In contrast, the Spin-Torque Ferromagnetic Resonance (ST-FMR) technique operates in the frequency domain[27-29]. By driving the magnetization into resonance with a radio-frequency current, ST-FMR amplifies the torque response, enabling highly sensitive and quantitative measurements of the effective spin Hall angle and allowing the disentanglement of different contributions to the total torque. ST-FMR not only overcomes sensitivity limitations but also provides a robust framework for analyzing spin and orbital transport contributions through frequency-domain analysis.

In this work, we employ the ST-FMR technique to investigate the spin and orbital Hall effects in a series of engineered metallic heterostructures. The samples investigated here consist of $SiO_2$/FM/NM trilayers deposited by sputtering, where FM denotes the ferromagnetic layer and NM the non-magnetic layer responsible for charge-to-spin or charge-to-orbital conversion. By comparing different NM materials, we analyze how their electronic and transport properties influence the resulting torques acting on the FM layer. Our results provide clear experimental evidence for orbital torque generation and demonstrate how ST-FMR serves as a powerful tool to decode the complex interplay between spin and orbital transport taking into account bulk and interface effects.

To ensure the article is self-contained and accessible, we begin with a theoretical review of the ST-FMR technique. Section II presents a didactic discussion of its fundamentals. Section III presents the results and discussion, and finally, the conclusions.

## II. Fundamentals of Spin-Torque Ferromagnetic Resonance

**Basic operating principle**

The ST-FMR technique relies on driving the magnetization of a ferromagnet/normal metal (FM/NM) bilayer into ferromagnetic resonance (FMR) by means of a radio-frequency (RF) electrical current. An RF current $I_{RF}$ is applied along the bilayer strip, where the NM layer typically possesses strong SOC. In the NM, this charge current is converted into an oscillating transverse spin current, primarily via the spin Hall effect, which is injected into the adjacent FM layer and exerts a spin–orbit torque on its magnetization $\vec{M}$. Simultaneously, the RF current flowing in the NM layer generates an oscillating Oersted magnetic field that acts as the main microwave excitation field for the magnetization. When the RF frequency matches the FMR condition, the combined action of this Oersted field and the spin–orbit torque drives the magnetization into resonant precession. The magnetization precession leads to a periodic modulation of the longitudinal resistance of the bilayer, dominated by the anisotropic magnetoresistance (AMR) of the FM layer. This time-dependent resistance $\Delta R(t)$ mixes with the applied RF current $I_{RF}(t)$, resulting in a rectified DC voltage, known as the mixing voltage $V_{\text{mix}}$, given by $V_{\text{mix}} = \langle I_{RF}(t)\, \Delta R(t) \rangle$, where $\langle ... \rangle$ denotes time averaging. Thus, $V_{\text{mix}}$ constitutes the primary experimental signal in ST-FMR measurements and is commonly used to extract information about spin-orbit torques. In addition, at resonance, the precessing magnetization pumps a spin current into the NM layer via the spin-pumping mechanism, which can be converted into a DC voltage through the inverse spin Hall effect. This contribution is typically much smaller than the AMR-induced mixing voltage and is therefore often neglected in the analysis of the main resonance line shape. The spin-pumping-induced voltage is typically orders of magnitude smaller than the AMR mixing signal and is therefore omitted in our analysis. It is also worth noting that the fraction of the RF current flowing directly through the FM layer generates an RF magnetic field whose in-plane components are symmetrically distributed with respect to the magnetization direction. As a result, the associated torques cancel by symmetry and do not contribute to the net excitation of the magnetization[27-33]. Fig. 1 illustrates the ST-FMR experimental geometry and electrical setup. On the left, an RF current flows along a SiO$_2$/FM/NM microstrip. Through spin-orbit coupling in the NM layer, a transverse spin current is generated, injecting damping-like and field-like spin-orbit torques into the FM magnetization. An oscillating Oersted field from the RF current also acts on the magnetization, driving precession around the effective field. On the right, the RF excitation from a signal generator is delivered through a bias-tee. The resulting rectified DC mixing voltage, produced via mixing of the RF current with the magnetoresistive response, is measured through the DC port using a lock-in amplifier. This enables electrical detection of the ferromagnetic resonance.

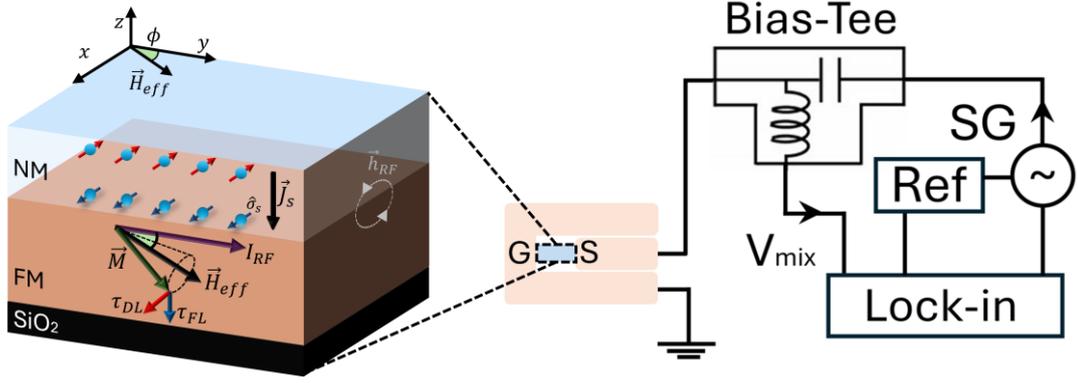

**FIG. 1**. ST-FMR setup. The $I_{RF}$ current that passes through the NM layer generates an oscillating spin current, which is responsible for the generation of an oscillating torque that acts on the magnetization $\vec{M}$. The mixing voltage $V_{mix}$ is detected using a lock-in amplifier and it is the combination of the AMR due to change in the FM layer resistance and the $\vec{h}_{RF}$ magnetic field induced by $I_{RF}$.

**Incorporating orbital torques and theoretical model**

In normal metals with weak SOC, the orbital Hall effect is expected to dominate over the spin Hall effect[34-37]. In this regime, an applied charge current predominantly generates a transverse orbital current through the intrinsic OHE[36,37]. However, the injection of a pure orbital current into an adjacent FM is not, by itself, sufficient to exert a torque on the magnetization. This is because the magnetic interaction requires spin angular momentum; therefore, a finite SOC in the FM is necessary to enable the conversion of orbital angular momentum into spin angular momentum and thus allow coupling to the FM magnetization[31,38]. When the NM possesses strong SOC, orbital-to-spin ($L \rightarrow S$) interconversion can already occur within the NM layer. Depending on the sign of the SOC correlation parameter $\langle \vec{L} \cdot \vec{S} \rangle$, the generated spin polarization may be parallel or antiparallel to the orbital polarization. In this case, a net spin current $\vec{J}_s$ is injected into the FM, directly exerting a spin-orbit torque on the magnetization $\vec{M}$. The direction of this torque is determined by the spin polarization $\hat{\sigma}_s$, whose orientation is governed by the sign and magnitude of $\langle \vec{L} \cdot \vec{S} \rangle$. In particular, for materials with $\langle \vec{L} \cdot \vec{S} \rangle < 0$, the spin polarization $\hat{\sigma}_s$ is antiparallel to the orbital polarization $\hat{\sigma}_L$. An alternative scenario occurs when the orbital current is injected into the FM and the $L \rightarrow S$ conversion occurs predominantly inside the FM due to its intrinsic SOC, thereby enabling interaction with the local magnetization. As a result, the effective torque acting on the FM magnetization reflects the combined contribution of the $L \rightarrow S$ conversion processes occurring in both layers, with their relative weights determined by the SOC correlations $\langle \vec{L} \cdot \vec{S} \rangle$ of the NM and FM. In this work, the SOC correlation parameter of the FM layer, $\langle \vec{L} \cdot \vec{S} \rangle_{FM}$, is taken to be positive for both ferromagnets studied, $Ni_{81}Fe_{19}$ (Permalloy - Py) and Ni[16]. In contrast, $\langle \vec{L} \cdot \vec{S} \rangle_{NM}$ depends on the specific NM material and is determined by the sign of its intrinsic spin and orbital conductivities[39]. Although recent studies have reported the presence of self-induced torques in Py and Ni[40], our measurements indicate damping-like torque efficiencies in these single-layer FMs that are orders of magnitude smaller than the orbital torque efficiencies observed in the corresponding FM/NM bilayers. It should be noted that while the self-torque measured in the standalone Ni(10) layer is negligible, its magnitude could, in principle, be modified by the presence of different interfaces due to changes in interfacial spin-orbit fields or orbital hybridization. In the heterostructures investigated here, however, any such interface-dependent self-torque contribution remains small compared to the bilayer-induced torques, in accordance with the refs [41,42]. All samples investigated in this work were fabricated by DC sputtering as $SiO_2$/FM($t_{FM}$)/NM($t_{FM}$) bilayers and were characterized without a capping layer. The ferromagnetic layer thicknesses were fixed at $t_{Py} = 5$ nm and $t_{Ni} = 10$ nm. The thickness of the NM layers was fixed at 8 nm, except for $CuO_x$ and Sb, which were 3 and 4 nm thick, respectively.

The magnetization dynamics under ST-FMR excitation is described by the Landau-Lifshitz-Gilbert (LLG) equation augmented with spin-torque terms[1, 27-33, 43-48]:

$$\frac{d\hat{m}}{dt} = -\gamma \hat{m} \times \vec{H}_{eff} + \alpha_G \left(\hat{m} \times \frac{d\hat{m}}{dt}\right) + \frac{\gamma \tau_{FL} \hat{m} \times \hat{\zeta}}{M_s} + \frac{\gamma \tau_{DL} \hat{m} \times (\hat{m} \times \hat{\zeta})}{M_s}. \quad (1)$$

Here, $\hat{m} = \vec{M}/M_s$ is the unit magnetization vector, $\gamma$ is the gyromagnetic ratio, $\alpha_G$ is the gilbert relaxation parameter, and $\vec{H}_{eff}$ is the effective field. The torque is decomposed into a field-like torque (FL) component $\tau_{FL}$ and a damping-like component $\tau_{DL}$, with $\hat{\zeta}$ representing the direction of the injected spin polarization. The corresponding effective fields are $H_{DL} = \tau_{DL}/M_s$ and $H_{FL} = \tau_{FL}/M_s$. Solving Eq. (1) for a system with Zeeman interaction and shape anisotropy, leads to an expression for the mixing voltage. For an in-plane magnetized film with the RF current applied along $x$ and an external field $\vec{H}_{ext}$ applied at an angle $\phi$ relative to the current, the signal takes the form [27-33, 43-48]:

$$V_{mix} = \frac{i_{RF}}{2\alpha_G \omega^+}\left(\frac{dR}{d\phi}\right)\left[\tau_{DL}(\phi) F_s(H) + \frac{\omega_2}{\omega}\tau_{FL}(\phi) F_A(H)\right]. \quad (2)$$

Where, $F_S(H) = \Delta H^2/[(H - H_r)^2 + \Delta H^2]$ and $F_A(H) = \Delta H(H - H_r)/[(H - H_r)^2 + \Delta H^2]$ represents the symmetric and antisymmetric Lorentzian components of $V_{mix}$, respectively, as illustrated in Fig. 2. Meanwhile, $R$ is the bilayer resistance, $\phi$ is the angle between external magnetic field $\vec{H}_{ext}$ and the direction of $I_{RF}$, $\Delta H = \alpha_G \omega/\gamma$ is the FMR linewidth, and the defined frequency-like quantities are $\omega = \gamma\sqrt{H(H + 4\pi M_{eff})}$, $\omega^+ = \gamma(2H_r + 4\pi M_{eff})$ and $\omega_2 = \gamma(H_r + 4\pi M_{eff})$, where $4\pi M_{eff}$ is the effective magnetization. Since the resistance change due to AMR is given by $\Delta R_{AMR}\cos^2\phi$, the symmetric and antisymmetric components are given by[27-29, 43-48]:

$$V_S = -\frac{i_{RF}}{2\alpha_G\omega^+}\Delta R_{AMR}\tau_{DL}^0 \sin 2\phi \cos\phi, \qquad (3.1)$$

$$V_A = -\frac{i_{RF}}{2\alpha_G\omega^+}\Delta R_{AMR}\tau_{FL}^0 \frac{\omega_2}{\omega}\sin 2\phi \cos\phi. \qquad (3.2)$$

Here, $\tau_{DL}^0$ and $\tau_{FL}^0$ are the amplitudes of the DL and FL torques. The signals are maximized when $\phi = 45º$ or $\phi = 225º$. Moreover, a positive symmetric signal $V_S$, implies a positive damping like torque $\tau_{DL}^0$, which implies a positive Hall spin angle $\theta_{SH}$, as exemplified in Fig. 2 for SiO$_2$/Py/Pt. To determine $\theta_{SH}$, it is necessary to calculate what is known as the torques efficiencies ($\xi_{DL}, \xi_{FL}$), that are defined by $\xi_{DL(FL)} = (2e/\hbar)(t_{FM}\tau_{DL(FL)}^0/j_c)$, where $j_c$ is the charge current density flowing in the NM and $t_{FM}$ is the FM layer thickness. In this work, we report two ways of calculating ($\xi_{DL}, \xi_{FL}$) based on previous works[27, 29-32, 49].

Fig. 2 represents the ST-FMR signals measured for SiO$_2$/FM/NM heterostrutures, where FM = Py or Ni and NM = Sb or Pt. The black symbols represent the raw mixing voltage ($V_{mix}$) obtained using the experimental setup shown in Fig. 1. In panel (a), the ST-FMR spectrum for the SiO$_2$/Py(5)/Sb(4) sample is shown. The blue curve corresponds to the fit obtained using Eq. (2), while the yellow and red curves represent the symmetric ($V_S$) and antisymmetric ($V_A$) components of $V_{mix}$, respectively. For this sample, the symmetric component is negative, and the $V_S/V_A$ ratio is relatively small. When the ferromagnetic layer is changed to Ni(10) in panel (b), the symmetric component reverses polarity and the $V_S/V_A$ ratio increases significantly. These changes reflect differences in the microscopic origin of the measured signal, which arises from charge-to-spin and charge-to-orbital conversion processes, SHE and OHE, respectively. Given that Ni exhibits stronger spin-orbit coupling than Py, orbital-related effects are expected to play a more significant role in Ni-based devices. The damping-like ($\xi_{DL}$) torque efficiency depends on factors such as FM layer thickness, current density in the NM layer, and resonance linewidth. Therefore, the Ni(10)/Sb(4) sample is anticipated to show a higher DL torque efficiency than Py(5)/Sb(4), consistent with the larger symmetric component observed in panel (b) relative to panel (a). To explore this behavior, control samples with Pt as the NM layer were also measured. Panels (c) and (d) display the ST-FMR spectra for SiO$_2$/Py(5)/Pt(8) and SiO$_2$/Ni(10)/Pt(8), respectively. In these samples, replacing Py with Ni again results in an increased $V_S/V_A$ ratio. Notably, the symmetric components in both Pt-based devices exhibit a positive polarity, consistent with the combined positive contributions expected from the SHE and OHE in Pt.

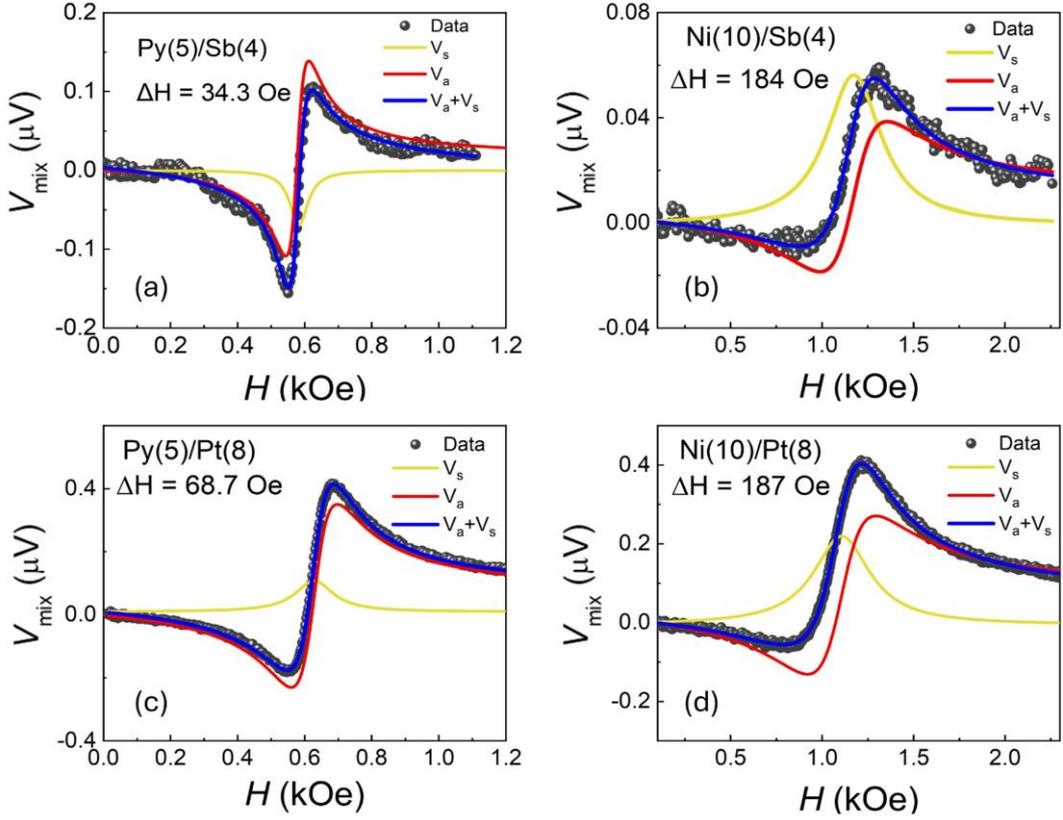

**FIG. 2.** ST-FMR spectra for SiO$_2$/FM/NM heterostructures. The black symbols represent the measured mixing voltage ($V_{mix}$) obtained from the setup shown in Fig. 1. The blue curves are fits to the data using Eq. (2), with the yellow and red lines showing the extracted symmetric ($V_S$) and antisymmetric ($V_A$) components, respectively. (a) Data for the Py(5)/Sb(4) sample, showing a small, negative symmetric component. (b) Data for the Ni(10)/Sb(4) sample, where the symmetric component reverses polarity and the $V_S/V_A$ ratio increases significantly. (c) Data for the Py(5)/Pt(8) sample. (d) Data for the Ni(10)/Pt(8) sample. Replacing Py with Ni in the Pt-based samples also leads to an increase in the $V_S/V_A$ ratio, with the symmetric components in (c) and (d) exhibiting a positive polarity. Comparing right panel with left panels, the difference in the increase of the symmetrical component is due to the presence of significant orbital torque in Ni, which has a strong SOC compared to Py. The data presented in this figure are original and were obtained in this work; no material from previously published sources is reproduced.

**Extraction of torque efficiencies**

The primary figure of merit is the damping-like torque efficiency $\xi_{\text{DL}}$, related to the effective spin/orbital Hall angle. Two common methods are used to extract it from ST-FMR data: (i) Amplitude ratio method and (ii) Direct angular fit method.

(i) Amplitude ratio method. For systems where Eqs. (3.1, 3.2) fully describe the angular dependence, one can use the ratio $V_S/V_A$. The torque efficiency can be obtained by writing an intermediate measurable efficiency $\xi_{\text{FMR}}$, which is directly calculated from the measured voltage ratio ($V_S/V_A$) and magnetic and geometric sample properties as:

$$\xi_{\text{FMR}} = \frac{V_S}{V_A} \frac{e\mu_0 M_s t_{\text{NM}} t_{\text{FM}}}{\hbar}\left[1 + \left(\frac{4\pi M_{eff}}{H_r}\right)\right]^{\frac{1}{2}}. \quad (4)$$

The measurable efficiency $\xi_{\text{FMR}}$ can be related, but not equal to, the fundamental $\xi_{\text{DL}}$, as:

$$\frac{1}{\xi_{\text{FMR}}} = \frac{1}{\xi_{\text{DL}}}\left(1 + \frac{\hbar}{e}\frac{\xi_{\text{FL}}}{\mu_0 M_s t_{\text{NM}} t_{\text{FM}}}\right). \tag{5}$$

Plotting $1/\xi_{\text{FMR}}$ for different $t_{\text{FM}}$ allows extraction of $\xi_{\text{DL}}$ and $\xi_{\text{FL}}$ from a linear fit. This first method of finding torque efficiencies is conventionally used when $V_S$ and $V_A$ only obey equations 3.1 and 3.2. This method is valid only when the angular dependence follows Eqs. (3.1, 3.2); deviations due to out-of-plane torques require the more general approach of this method.

(ii) Direct angular fit method. In many cases, especially when unconventional out-of-plane torque components $(\tau_{\text{DL}}^{\perp}, \tau_{\text{FL}}^{\perp})$ are present, the angular dependence deviates from the simple $\sin 2\phi \cos \phi$ form. A more robust approach is to measure $V_S$ and $V_A$ as a function of $\phi$ and fit the full angular dependence:

$$V_S(\phi) \propto \left(\tau_{\text{DL}}^0 \cos\phi + \tau_{\text{DL}}^{\perp}\right)\sin 2\phi, \tag{6}$$

$$V_A(\phi) \propto \left(\tau_{\text{FL}}^0 \cos\phi + \tau_{\text{FL}}^{\perp}\right)\sin 2\phi. \tag{7}$$

Where the torques $\tau_{\text{DL}}^0$, and $\tau_{\text{FL}}^0$ for $SiO_2$/FM/NM are expressed as:

$$\tau_{\text{DL}}^0 = -\frac{2\alpha_G \omega^+}{i_{RF}\Delta R_{AMR}} A_S, \tag{8}$$

$$\tau_{\text{FL}}^0 = -\frac{2\alpha_G \omega^+}{i_{RF}\Delta R_{AMR}}\frac{\omega}{\omega_2} B_A. \tag{9}$$

Here, $A_S$ and $B_A$ is the amplitude of $\sin 2\phi \cos \phi$ term. The torque efficiency $\xi_{\text{DL}}$ becomes $\xi_{\text{DL}} = \tau_{\text{DL}}^0 e M_s t_{\text{FM}}/\mu_B j_{\text{NM}}$. In some measurements of $V_S$ and $V_A$ as a function of $\phi$, we observed unconventional behaviors that have already been reported in several experimental works[50-52]. The angular dependence can be well interpreted when we add the extra term $\sin 2\phi$, which arises from the presence of out-of-plane torques $\tau_{\text{DL}}^{\perp}$ and $\tau_{\text{FL}}^{\perp}$, associated with spin or orbital z-polarizations at the FM/NM interface due to inversion symmetry breaking. Thus, the torques must then be $\tau_{\text{DL}}(\phi) = \tau_{\text{DL}}^0 \cos\phi + \tau_{\text{DL}}^{\perp}$, and $\tau_{\text{FL}}(\phi) = \tau_{\text{FL}}^0 \cos\phi + \tau_{\text{FL}}^{\perp}$ (see Table I), as expressed by equations (6) and (7). In this context, we separate the contributions of conventional torques (Hall effects, Rashba effects) from the contributions of out-of-plane torques. The observation of out-of-plane torque components indicates additional symmetry breaking beyond that expected for ideal polycrystalline metallic bilayers. This symmetry reduction likely originates at the FM/NM interface, where structural inversion asymmetry, strain, or interfacial disorder can generate angular momentum with out-of-plane polarization. As seen in Table I, no simple correlation with the specific FM or NM material is observed, suggesting that these torques are mainly governed by interface-specific effects rather than bulk transport properties.

In this work, we treat the FM/NM system as two parallel resistors with layer resistivities independent of the applied frequency and with bulk material values provided in the literature (see Table II). The current density $j_{NM}$ passing through the NM layer can be easily determined by $j_{\text{NM}} = I_{RF}/\left[wt_{\text{NM}}\left(1 + \frac{\rho_{\text{NM}} t_{\text{FM}}}{\rho_{\text{FM}} t_{\text{NM}}}\right)\right]$, where $w$ is the width of the strip, $\rho_{\text{FM}}$ and $\rho_{\text{NM}}$ are the resistivities of the FM and NM layers, respectively. The saturation magnetization $M_s$ for the two ferromagnetic materials used in this work was $M_s^{Py} = 8.0 \times 10^5$ A/m and $M_s^{Ni} = 4.5 \times 10^5$ A/m, for Py and Ni, respectively.

Finally, a brief comment on the analysis scheme. In our ST-FMR evaluation, we minimized possible artifacts, such as spin pumping, self-torque, and heating, by using moderate RF power and verifying signal linearity. The main difficulty instead stemmed from additional torque contributions, revealed by the angular dependence of the symmetric-to-antisymmetric ratio ($V_S/V_A$). Because this behavior violates the single-symmetry assumption underlying conventional linear fits, such methods become unreliable. We therefore adopted a direct calculation of the $\xi_{\text{DL}}$. Although this procedure requires precise RF current calibration, it allows us to disentangle standard in-plane torques from unconventional out-of-plane components, ensuring robust torque quantification in the presence of multiple symmetries.

**III. RESULTS AND DISCUSSION**

Fig. 3 shows the $V_{\text{mix}}$ signals for the SiO$_2$/Py/Pt bilayers, highlighting the frequency and angular dependencies. To start with the torques evaluation, we will show that the two methods of calculating the torques efficiencies presented in the introduction should give similar results. Therefore, let's start with the first approach, which is the evaluation of the torques efficiencies by varying the thickness of the FM layer in the samples SiO$_2$/Py($t_{\text{FM}}$)/Pt(8), with $t_{\text{FM}} = 3, 4, 5$ and 6 nm, the number in parentheses indicates the film thickness.

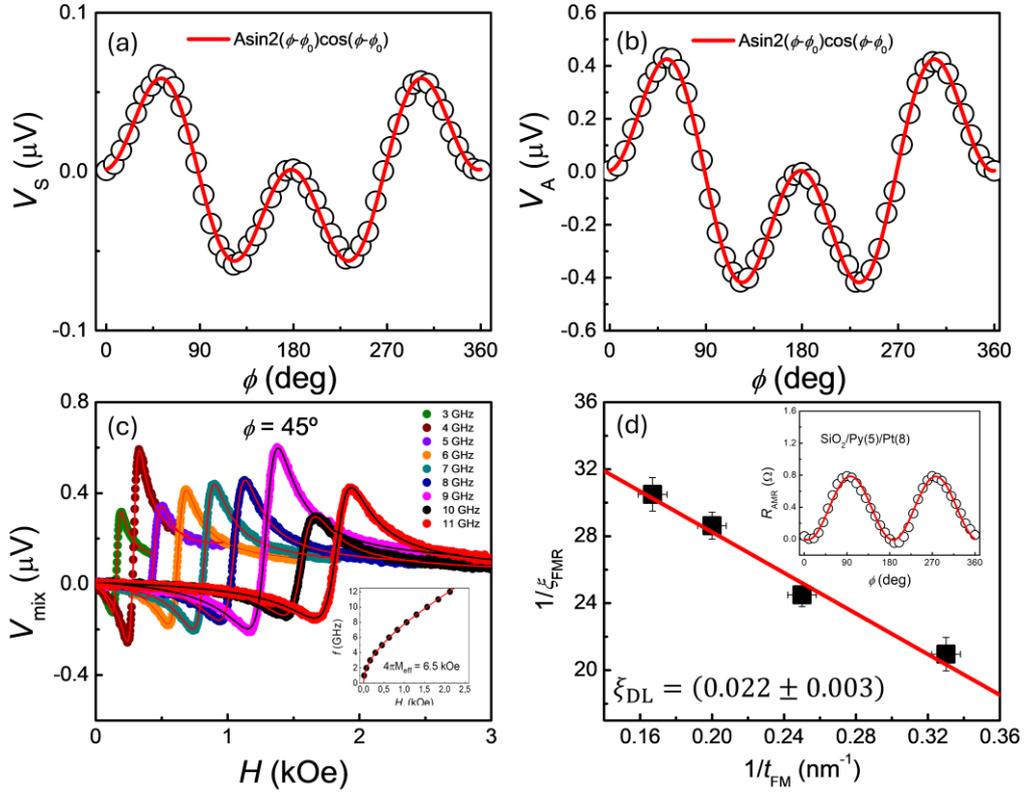

**FIG. 3**. (a,b) ST-FMR measurements of the symmetrical $V_S$ and antisymmetrical $V_A$ parts as a function of the $\phi$ angle, resulting in typical behavior for this technique. (c) ST-FMR curves for Py(5)/Pt(8) as a function of RF frequency. The inset shows the Kittel curve for Py(5)/Pt(8). (d) $1/\xi_{FMR}$ versus $1/t_{FM}$ for Py($t_{FM}$)/Pt(8), where $t_{FM} = 3, 4, 5\ e\ 6$ nm. The RF power was 8 mW and the microwave frequency was 6 GHz, except for the results shown in (c). The inset shows the anisotropic magnetoresistance (AMR) relative to the bilayer SiO$_2$/Py(5)/Pt(8).

Fig. 3 (a,b) shows the angular dependence of $V_S$ and $V_A$ as a function of $\phi$. From Eq. 4, we take the ratio between the symmetric and antisymmetric components $V_S/V_A$ shown in Fig. 3 (a), preferentially choosing a specific angle $\phi = 45°$ (Fig. 3(c)). We also obtain the factor $4\pi M_{eff}$ by varying the frequency of $I_{RF}$ and fitting the resulting curve by the Kittel equation $\omega = \gamma\sqrt{H(H + 4\pi M_{eff})}$ (inset Fig. 3(c)). Using these data, the linear fit of the effective torque efficiency as a function of the inverse thickness $t_{FM}$ can be performed, as displayed in Fig. 3 (d). Therefore, we obtain a damping like torque efficiency $\xi_{DL} = 0.022 \pm 0.003$. In contrast, using the second approach, a value of $\xi_{DL} = (0.013 \pm 0.003)$ was obtained, which is somewhat smaller than the value obtained from the first method, but still of the same order of magnitude, indicating qualitative agreement between the two approaches. As this work investigated several samples, where in some of them we observed torque terms out of the plane, it was necessary to calculate the torque $\xi_{DL}$ directly. Fig. 4 shows the dependence of $V_S$ and $V_A$ as a function of $\phi$ for SiO$_2$/FM/NM, FM being Ni or Py and NM being Ag or Bi, which differs from the results presented in Fig. 3 (a,b), requiring the addition of an extra term $\sin 2\phi$ for the fit in red. Appendix A presents the experimental results obtained for $V_S$ and $V_A$ for all samples measured in this work, while Appendix B presents the AMR measurements in the same samples.

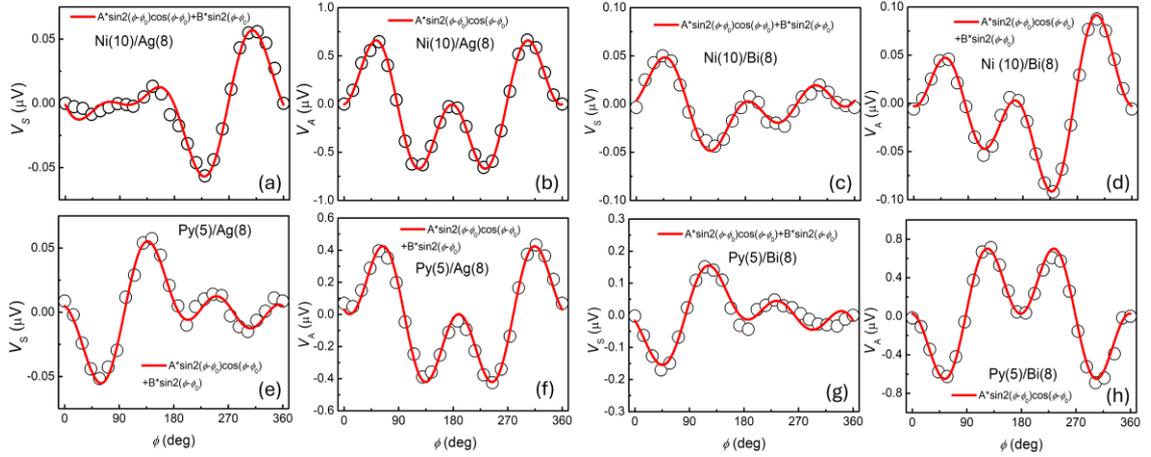

**FIG. 4.** (a-h) Angular dependence of the symmetric ($V_S$) and antisymmetric ($V_A$) components obtained from ST-FMR. The sample composition is indicated in each panel. The solid lines represent fits to the data including an additional odd-in-$\phi$ contribution of the form $\sin 2(\phi - \phi_0)$, which is required to accurately reproduce the observed angular dependence in all cases.

Unconventional spin or orbital polarizations (e.g., out-of-plane components) can, in principle, contribute to the angular asymmetry observed in $V_{\text{mix}}$. For example, a $z$-polarized orbital current, arising from symmetry breaking at the interface, would interact differently with the precessing magnetization, thereby generating additional torque components that modify the angular dependence. Although such phenomena may be at play, unraveling them requires careful analysis of the materials involved. In our specific case, the contrasting behavior between Py/Bi and Ni/Bi, where both exhibit clear damping-like torques, with Ni/Bi showing a larger magnitude than Py/Bi (see, Fig. 5), allows us to draw a more specific conclusion. Since polycrystalline Bi exhibits negligible bulk spin-charge conversion[53,54], the observed torques must originate from an interface-driven mechanism. The stronger response in Ni/Bi, where Ni is known to efficiently generate orbital currents, points to a significant orbital contribution at the Bi interface, complementing the spin-mediated channel. This behavior contrasts with Ni/Pt, where torques are predominantly spin-mediated due to bulk spin Hall effect of Pt. Our results thus establish that torque generation in Bi bilayers arises from a mixed spin+orbital origin, with the dominant channel determined by the angular momentum character of the adjacent ferromagnet. In addition to analyzing the contributions of DL and FL torque in the Ni/Bi and Py/Bi samples, we also investigated the influence of the FM/Bi interface on the observed ST-FMR signals (see Appendix C for more details). However, a detailed and systematic investigation of such phenomena in each material is beyond the scope of this work, since it would require higher-resolution angular sampling and complementary measurement techniques. Further work is necessary to fully clarify and describe the complete physical picture.

Fig. 5 presents the values obtained for $\xi_{\text{DL}}$, calculated from Eq. 6. We chose several NM materials with strong OHE[34-37], such as Zr, Mo, Ta, Ir, and also interfaces with $CuO_x$[49,55-57]. The results presented in Fig. 5 highlight the relevance of orbital-related mechanisms in the generation of damping-like torque efficiencies, $\xi_{\text{DL}}$, across different FM/NM bilayers. A clear distinction emerges when comparing systems based on Py and Ni as ferromagnetic layers. SiO$_2$/Ni(10)/NM consistently exhibits significantly larger $\xi_{DL}$ values than SiO$_2$/Py(5)/NM for several normal metals, which can be attributed primarily to the intrinsically stronger SOC of Ni[39]. This enhanced SOC in Ni facilitates a more efficient intra-layer orbital-to-spin conversion within the ferromagnetic layer, thereby amplifying the resulting torque. Moreover, the observed dependence of $\xi_{DL}$ cannot be explained solely by conventional spin Hall effects. Different normal metals are known to exhibit distinct magnitudes and even signs of both spin Hall conductivity and orbital Hall conductivity[34-36], depending on their electronic structure and symmetry. In this context, materials such as Pt, Bi, and Sb display strong positive orbital torque contributions when interfaced with Ni[54,58], consistent with a sizable orbital Hall effect in the NM layer, followed by efficient orbital-to-spin conversion at the interface or inside the ferromagnet. In contrast, materials such as Ag exhibit reduced efficiencies, likely due to its weak spin-orbit coupling and negligible orbital Hall conductivity. The comparison between Py/NM and Ni/NM bilayers further reinforces the orbital nature of the observed torques. While Py, characterized by relatively weak SOC, shows modest torque efficiencies dominated by conventional spin-current-driven mechanisms, Ni demonstrates a pronounced sensitivity to the orbital current injected from the NM layer. This behavior provides strong experimental evidence that orbital currents, generated via the orbital Hall effect in the normal metal, play a dominant role in the torque generation when the ferromagnetic layer possesses sufficiently strong SOC to mediate orbital-to-spin angular momentum conversion.

A particularly remarkable result is the magnitude of the damping-like torque efficiency, $\xi_{DL}$, at interfaces involving CuO$_x$. We observe large values for the Ni/CuO$_x$ bilayer, whereas the efficiency is several times smaller in Py/CuO$_x$, indicating a strong contribution from orbital effects, which are already well established in CuO$_x$[55-57]. In this case, the dominant mechanism is associated with the orbital Rashba effect in CuO$_x$, which efficiently generates an orbital current at the interface. On the other hand, although interfacial contributions cannot be completely ruled out in the other FM/NM bilayers, our results indicate that both bulk and interfacial effects in the various materials studied are predominantly orbital in nature. These findings provide important insights into the study and control of orbital torques in magnetic heterostructures, further advancing the understanding of the underlying physical mechanisms. In contrast, for Zr[59] and Mo we observe torque efficiencies significantly lower than theoretically predicted, despite their expected strong orbital Hall effect. In particular, the negative damping-like torque observed in the Ni/Mo system presents an intriguing contrast with the positive orbital Hall conductivity extracted from inverse orbital Hall effect measurements in similar materials. This apparent sign discrepancy does not represent a contradiction, but rather highlights the complex nature of orbital transport in thin-film heterostructures. Recent theoretical work suggests that non-reciprocity between direct and inverse orbital-charge conversion processes can occur due to the non-conserved nature of orbital angular momentum[60]. Additionally, competing contributions from the bulk orbital Hall effect and interfacial conversion mechanisms, sensitive to factors such as lattice mismatch and electronic structure, could ultimately determine the sign of the net torque. This observation underscores that the direct and inverse orbital Hall effects are not simply reciprocal phenomena in complex heterostructures, and their relationship warrants further systematic investigation. Overall, these results demonstrate that both the intrinsic SOC of the ferromagnetic layer and the orbital Hall conductivity of the normal metal are key parameters governing the efficiency of damping-like torques. This establishes orbital torque as a robust and material-dependent mechanism, expanding the conventional spin Hall paradigm and opening new pathways for magnetization control using orbital angular momentum.

Finally, to evaluate possible thermal contributions, we performed rf-power-dependent ST-FMR measurements on a control sample of $SiO_2/Py(5)/Pt(8)$. The resonance linewidth remains essentially constant with power, while the symmetric ST-FMR amplitude scales linearly up to ~10 mW. As all measurements in this work were performed at 8 mW, these results confirm operation in the linear regime and indicate that rf-induced heating or spin-Seebeck-related backgrounds do not significantly affect the extracted torque parameters (see appendix D).

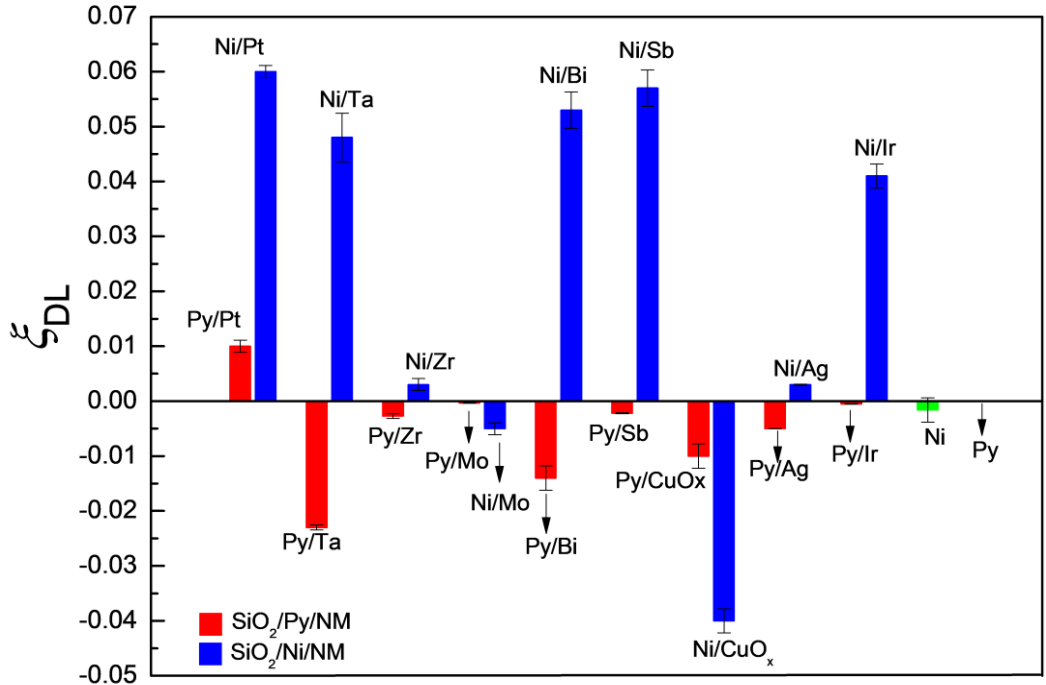

**FIG. 5.** Main results for damping-like efficiency $\xi_{DL}$ in $SiO_2/FM/NM$ for an RF frequency of 6 GHz and applied power of 8 mW. The thicknesses of the ferromagnetic layers (FM) were 5 nm for Py and 10 nm for Ni. The nonmagnetic (NM) layer had a standard thickness of 8 nm, except for Sb (4 nm) and $CuO_x$ (3 nm). With the exception of $CuO_x$, which was already expected to exhibit a strong ORE[22], the other NM layers did not show surface oxidation effects that significantly influenced the overall results. The $\xi_{DL}$, is higher in samples with Ni than in those with Py, which can be attributed to the stronger spin-orbit coupling in Ni compared to Py. As a result, orbital effects become more significant, enhancing the magnitude of $\xi_{DL}$.

## IV. CONCLUSION

In this work, we experimentally investigated spin and orbital torque phenomena in FM/NM heterostructures using the ST-FMR technique. By employing Py and Ni as ferromagnetic layers, we were able to distinguish between torque mechanisms dominated by the SHE and those primarily governed by the OHE. Our results reveal that Ni/NM bilayers exhibit significantly larger damping-like torque efficiencies $\xi_{DL}$ compared to Py/NM systems. This behavior highlights the crucial role of orbital angular momentum and the strong intrinsic SOC in Ni, which enhances the orbital-to-spin conversion and amplifies the generated torque. The systematic comparison across different normal metals demonstrates that both the magnitude and sign of $\xi_{DL}$ are closely linked to the orbital Hall conductivity of each material, confirming the predominantly orbital nature of the observed torques. Materials such as Bi, Sb, and $CuO_x$ interface display significant orbital-driven torque efficiencies, while Zr and Mo show smaller-than-expected efficiencies despite theoretical predictions of strong OHE. This deviation is attributed to interface sensitivity and surface oxidation effects, emphasizing that interfacial chemistry and environmental exposure critically affect orbital torque generation. Overall, our findings demonstrate that the damping-like torque efficiency is jointly determined by the orbital Hall conductivity of the normal metal and the SOC strength of the ferromagnet. These results establish orbital torque as a robust, material-dependent mechanism that extends the conventional spin Hall paradigm and opens new opportunities for magnetization control through orbital angular momentum in spintronic and orbitronic devices. Future work may focus on engineering interfaces to enhance orbital transmission and mitigate oxidation effects, paving the way for efficient orbital-torque-based memory and logic devices.

**ACKNOWLEDGMENTS**

This research is supported by Conselho Nacional de Desenvolvimento Científico e Tecnológico (CNPq), Coordenação de Aperfeiçoamento de Pessoal de Nível Superior (CAPES) (Grant No. 0041/2022), Financiadora de Estudos e Projetos (FINEP), Fundação de Amparo à Ciência e Tecnologia do Estado de Pernambuco (FACEPE) – (Grant No. BFP-0345 1.05/24), Universidade Federal de Pernambuco, Multiuser Laboratory Facilities of DF-UFPE, Fundação de Amparo à Pesquisa do Estado de Minas Gerais (FAPEMIG) - Rede de Pesquisa em Materiais 2D and Rede de Nanomagnetismo, and INCT of Spintronics and Advanced Magnetic Nanostructures (INCT-SpinNanoMag), CNPq 406836/2022-1.

**AUTHOR DECLARATIONS**
**Conflict of Interest**
The authors have no conflicts to disclose.

**TABLE I.** Damping-like ($\tau_{DL}$) and field-like ($\tau_{FL}$) torques were obtained for the SiO$_2$/FM/NM bilayers used in this work. In-plane and out-of-plane torques were obtained, where the latter originates from spin-orbital $z$-polarizations, which can dynamically affect $\tau_{DL}$ and $\tau_{FL}$ torques.

| NM | $\tau_{DL}$[Py/NM] $(10^5 s^{-1})$ | $\tau_{DL}$[Ni/NM] $(10^5 s^{-1})$ | $\tau_{FL}$[Py/NM] $(10^5 s^{-1})$ | $\tau_{FL}$[Ni/NM] $(10^5 s^{-1})$ |
|---|---|---|---|---|
| Zr | $\tau_{DL}^0 = -0.29$ | $\tau_{DL}^0 = -0.17$ $\tau_{DL}^\perp = 1.21$ | $\tau_{FL}^0 = 0.66$ | $\tau_{FL}^0 = 0.28$ $\tau_{FL}^\perp = -0.22$ |
| Ir | $\tau_{DL}^0 = -0.20$ $\tau_{DL}^\perp = 0.27$ | $\tau_{DL}^0 = 9.97$ | $\tau_{FL}^0 = 7.70$ | $\tau_{FL}^0 = 14.80$ |
| Mo | $\tau_{DL}^0 = -0.10$ | $\tau_{DL}^0 = -0.67$ | $\tau_{FL}^0 = 1.50$ | $\tau_{FL}^0 = -0.61$ |
| Pt | $\tau_{DL}^0 = 3.52$ | $\tau_{DL}^0 = 12.70$ | $\tau_{FL}^0 = 4.70$ | $\tau_{FL}^0 = 11.01$ |
| Ta | $\tau_{DL}^0 = -3.67$ | $\tau_{DL}^0 = 4.38$ $\tau_{DL}^\perp = -0.57$ | $\tau_{FL}^0 = 1.60$ | $\tau_{FL}^0 = 1.33$ $\tau_{FL}^\perp = -0.18$ |
| Sb | $\tau_{DL}^0 = -0.12$ | $\tau_{DL}^0 = 4.10$ $\tau_{DL}^\perp = 0.43$ | $\tau_{FL}^0 = -1.10$ $\tau_{FL}^\perp = 0.10$ | $\tau_{FL}^0 = 9.98$ |
| Bi | $\tau_{DL}^0 = -0.73$ $\tau_{DL}^\perp = -0.34$ | $\tau_{DL}^0 = 0.90$ $\tau_{DL}^\perp = 0.30$ | $\tau_{FL}^0 = -1.30$ | $\tau_{FL}^0 = 0.93$ $\tau_{FL}^\perp = -0.23$ |
| CuOx | $\tau_{DL}^0 = -0,73$ | $\tau_{DL}^0 = -1.26$ | $\tau_{FL}^0 = -0.0055$ $\tau_{FL}^\perp = -0.037$ | $\tau_{FL}^0 = 0.045$ $\tau_{FL}^\perp = -0.015$ |
| Ag | $\tau_{DL}^0 = -4.10$ $\tau_{DL}^\perp = -2.20$ | $\tau_{DL}^0 = 1.96$ $\tau_{DL}^\perp = -1.74$ | $\tau_{FL}^0 = 17.20$ | $\tau_{FL}^0 = 19.1$ |
| Ni | $\tau_{DL}^0 = -0.56$ | $\tau_{FL}^\perp = -0.28$ | | |
| Py | $\tau_{DL}^0 = -0.026$ | $\tau_{FL}^\perp = -0.89$ | | |

**TABLE II.** Resistivity values for the thin films used in the $\xi_{DL}$ calculations.

| Material | $\rho(\mu\Omega.\,cm)$ |
|---|---|
| Py | 40 |
| Ni | 20 |
| Pt | 50 |
| Ta | 100 |
| Zr | 90 |
| Mo | 35 |
| Bi | 420 |
| Sb | 120 |
| Ag | 10 |
| Ir | 30 |
| CuO$_x$(3) | 225 |

# Appendix A

Fig. 6 shows the angular dependence of the symmetric $V_S$ (blue symbols) and antisymmetric $V_A$ (red symbols) components of the rectified voltage $V_{mix}$ as a function of the azimuthal angle in the $\phi$. Measurements were performed at a fixed RF frequency of 6 GHz and an applied RF power of 8 mW (effective value reaching the sample). Data are presented for different multilayer structures. For all samples, both $V_S(\phi)$ and $V_A(\phi)$ exhibit clear periodic modulations with $\phi$. The angular dependence can generally be described by combinations of $\sin(2\phi)\cos(2\phi)$. The antisymmetric component $V_A$ typically exhibits larger amplitudes, consistent with field-like torque contributions and Oersted field-induced effects. In contrast, the symmetrical $V_S$ voltage exhibits smaller, yet systematic, angular variations, which are commonly associated with damping-like torques and out-of-plane torque contributions, requiring the addition of the $\sin 2\phi$ term for proper fit. The $V_S$ and $V_A$ measurements obtained for SiO$_2$/Py and SiO$_2$/Ni (black symbols), representing the self-torque of the FM, are also presented.

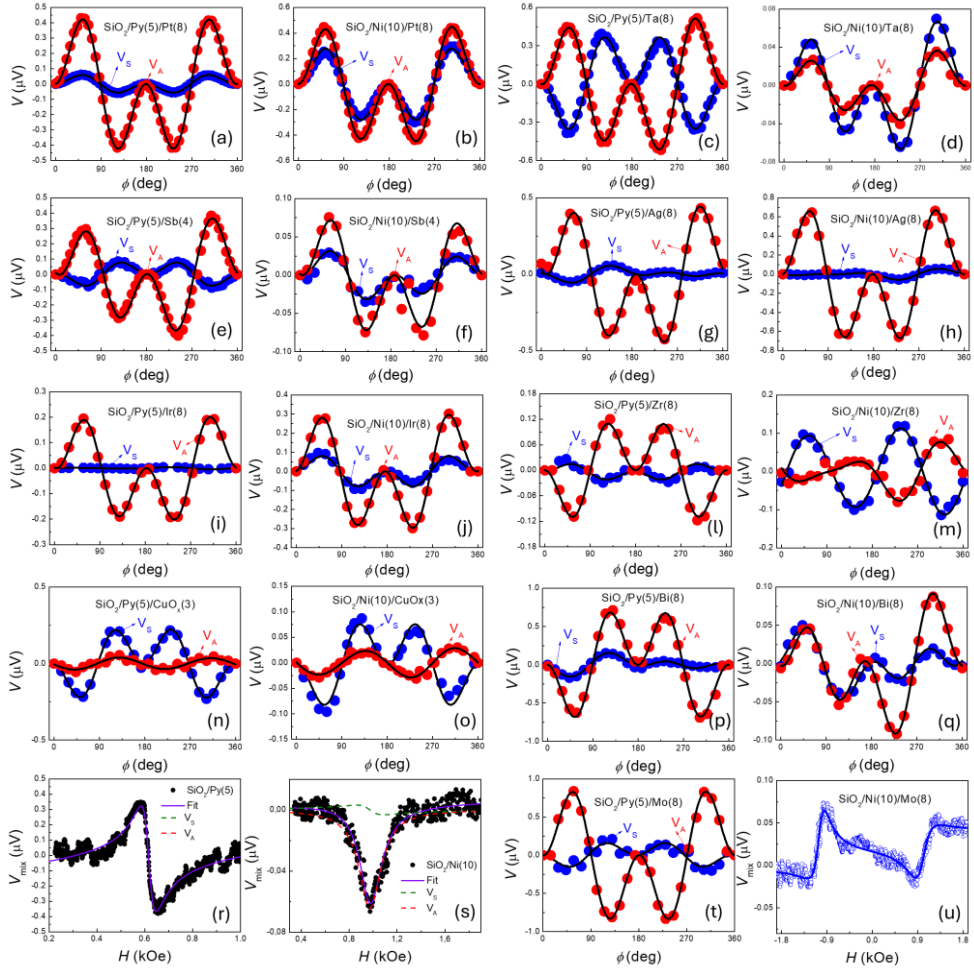

**FIG. 6.** (a-u) Angular dependence of $V_S$ (blue symbols) and $V_A$ (red symbols), obtained by ST-FMR, as a function of the azimuthal angle $\phi$. The measurements were performed for a fixed RF frequency of 6 GHz and an RF power of 8 mW, the effective value reaching the sample. The sample composition is indicated in each panel.

# Appendix B

Fig. 7 shows the angular dependence of the anisotropic magnetoresistance $R_{AMR}$ as a function of the in-plane azimuthal angle $\phi$ for the different multilayer stacks studied. In all samples, $R_{AMR}(\phi)$ exhibits a clear twofold symmetry with a 180° periodicity, characteristic of conventional AMR in ferromagnetic thin films, and is well described by a $\cos^2(\phi)$ dependence. The solid red lines correspond to fits using the standard AMR expression and show excellent agreement with the experimental data. Differences in the AMR amplitude among the various material combinations and thicknesses reflect variations in resistivity and interfacial scattering, while the robust angular behavior confirms reliable magnetization alignment during the ST-FMR measurements and supports the angular analysis used to extract $V_S$ and $V_A$.

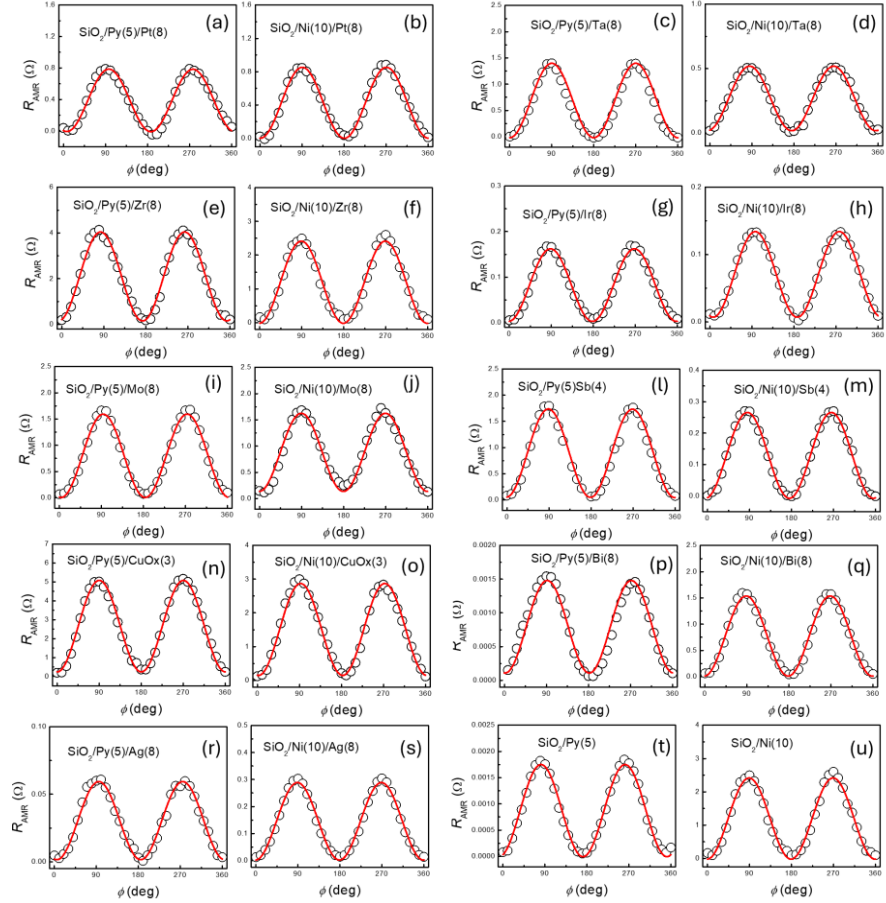

**FIG. 7.** (a-u) Anisotropic Magnetoresistance ($R_{AMR}$) for each bilayer. The data were obtained from azimuthal angular scanning, maintaining a fixed external magnetic field of 2 kOe and an applied electric current on the order of 1 mA. The sample composition is indicated in each panel.

# Appendix C

Fig. 8 shows the angular dependence of the symmetric ($V_S$) and antisymmetric ($V_A$) ST-FMR voltage components for (a) SiO$_2$/Py(5)/Bi(8), (b) SiO$_2$/Py(5)/Au(2)/Bi(8)/Au(1), (c) SiO$_2$/Ni(10)/Bi(8), and (d) SiO$_2$/Ni(10)/Au(2)/Bi(8)/Au(1). For the direct FM/Bi bilayers [(a) and (c)], the angular dependence exhibits clear asymmetries, particularly pronounced in the Ni/Bi sample, indicating the presence of interfacial Rashba-like contributions that require an additional odd-in-$\phi$ term for an adequate description. In contrast, the Py/Bi sample shows a dominant antisymmetric component, consistent with a strong contribution from Oersted fields or field-like torques, while still displaying a finite symmetric component associated with out-of-plane torques. Motivated by a recent study in which the insertion of a Cu layer between Cr and NiO layers suppressed orbital-to-spin conversion, we investigate analogous effects by inserting a thin layer of Au between the FM layer and Bi film[61]. Upon insertion of a thin Au spacer [(b) and (d)], the angular dependence of both $V_A$ and $V_S$ becomes markedly more symmetric for Py- and Ni-based systems. This behavior indicates that decoupling the FM from Bi suppresses interfacial Rashba-like effects. The overall increase in signal amplitude, particularly in $V_A$, further suggests that the Au layer modifies the current distribution and effective fields within the stack. These results demonstrate that the asymmetric contributions, and consequently the out-of-plane torque components, originate predominantly from the direct FM/Bi interface. Although Bi is used here as a model system, similar interfacial mechanisms are expected to be relevant for other nonmagnetic layers in contact with ferromagnets.

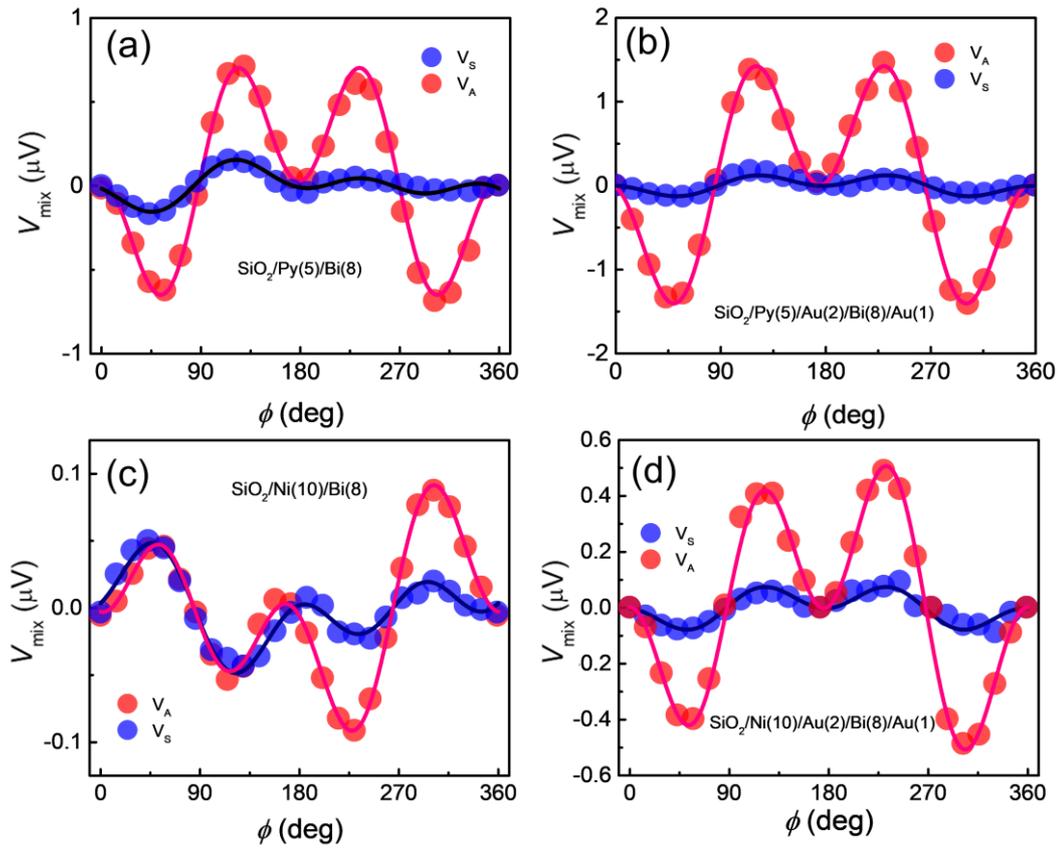

**FIG. 8.** Angular dependence of the mixing voltage $V_{mix}$ obtained from ST-FMR) measurements. The data are decomposed into symmetric ($V_S$, blue) and antisymmetric ($V_A$, red) components for the structures (a) $SiO_2$/Py/Bi, (b) $SiO_2$/Py/Au/Bi/Au, (c) $SiO_2$/Ni/Bi, and (d) $SiO_2$/Ni/Au/Bi/Au. Symbols represent experimental data and solid lines are fits. Direct FM/Bi bilayers [(a),(c)] exhibit pronounced asymmetries in the angular dependence, whereas the insertion of an Au spacer [(b),(d)] restores a largely symmetric behavior, indicating suppression of interfacial Rashba-like contributions.

# Appendix D

This appendix presents RF-power-dependent ST-FMR measurements performed on a control sample of $SiO_2/Py(5)/Pt(8)$ in order to evaluate possible thermal contributions to the detected voltage signal. Fig. 9(a) shows representative ST-FMR spectra measured at different microwave frequencies (6-10 GHz) for a fixed RF power of 8 mW, confirming the expected resonance line shape and frequency-dependent resonance field characteristic of ferromagnetic resonance in the Py layer. Fig. 9(b) displays the evolution of the ST-FMR signal as a function of RF power at a fixed frequency of 6 GHz and an in-plane field angle of $\phi = 45°$. The amplitude of the resonance signal increases systematically with increasing power while preserving the same line shape. The resonance linewidth $\Delta H$ extracted from the fits is plotted as a function of RF power in Fig. 9(c). Within the investigated power range, $\Delta H$ remains essentially constant, indicating that nonlinear effects such as RF-induced heating or spin-wave excitation are negligible. Fig. 9(d) shows the symmetric voltage component $V_S$ as a function of RF power. A linear dependence is observed up to approximately 10 mW, confirming that the measurements are performed within the linear response regime. Since all measurements reported in the main text were carried out at an RF power of 8 mW, these results demonstrate that the experiments were conducted within the linear regime and that possible thermal contributions, such as RF heating or spin-Seebeck-related effects, do not significantly affect the extracted torque parameters.

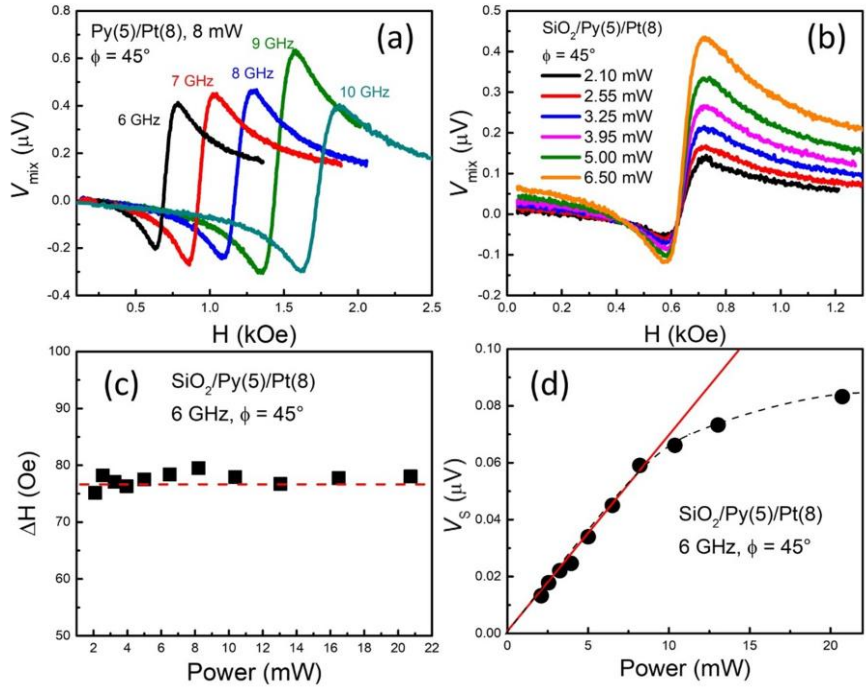

**FIG. 9.** RF-power dependence of ST-FMR measurements in a control sample of $SiO_2/Py(5)/Pt(8)$. (a) ST-FMR spectra measured at different microwave frequencies (6-10 GHz) at a fixed RF power of 8 mW. (b) Evolution of the ST-FMR signal with increasing RF power at 6 GHz and $\phi = 45°$. (c) Resonance linewidth $\Delta H$ as a function of RF power, showing no significant variation within the investigated range. (d) Symmetric voltage component $V_S$ as a function of RF power. The linear dependence observed up to ~10 mW confirms that the measurements are performed within the linear regime.

DOI: https://doi.org/10.1103/nns5-cn81